\documentclass[graybox, envcountchap, authoryear, natbib]{svmult}

\usepackage{mathptmx}	    
\usepackage{helvet}	    
\usepackage{courier}	    
\usepackage{type1cm}	    

\usepackage{makeidx}	     
\usepackage{graphicx}	     
\usepackage{multicol}	     
\usepackage[bottom]{footmisc}

\usepackage{natbib} 
\usepackage{amsmath} 
\usepackage{amssymb} 

\bibpunct{(}{)}{;}{a}{}{,}

\defcitealias{lehner13}{L13}
\defcitealias{lehner16}{L16}
\defcitealias{wotta16}{W16}
\defcitealias{fumagalli16}{FOP16}

\newcommand{\aj}{AJ}
\newcommand{\apj}{ApJ}
\newcommand{\apjl}{ApJL}     
\newcommand{\apjs}{ApJS}
\newcommand{\aap}{A\&A}
\newcommand{\aapr}{A\&A~Rv}
\newcommand{\mnras}{MNRAS}
\newcommand{\nat}{Nature}

\newcommand{\rmxaa}{RMxAA}

\newcommand{\xh}{{[\rm X/H]}}

\newcommand{\cmmm}{cm$^{-3}$}

\newcommand{\lya}{Ly$\alpha$}

\newcommand{\nhi}{$N_{\rm HI}$}
\newcommand{\nmetals}{$N_{\rm metals}$}

\newcommand{\novi}{$N_{\rm OVI}$}
\newcommand{\nhii}{$N_{\rm HII}$}
\newcommand{\mnhi}{N_{\rm HI}}

\newcommand{\mlnovi}{\log N_{\rm OVI}}

\newcommand{\mlnhi}{\log N_{\rm HI}}

\newcommand{\kms}{${\rm km\,s}^{-1}$}
\newcommand{\hst}{{\em HST}}

\newcommand{\la}{\lesssim}
\newcommand{\ga}{\gtrsim}

\newcommand{\hi}{H$\;${\small\rm I}\relax}

\newcommand{\cii}{C$\;${\small\rm II}\relax}
\newcommand{\ciii}{C$\;${\small\rm III}\relax}
\newcommand{\civ}{C$\;${\small\rm IV}\relax}

\newcommand{\oi}{O$\;${\small\rm I}\relax}

\newcommand{\oiii}{O$\;${\small\rm III}\relax}

\newcommand{\ovi}{O$\;${\small\rm VI}\relax}

\newcommand{\siii}{Si$\;${\small\rm II}\relax}
\newcommand{\siiii}{Si$\;${\small\rm III}\relax}

\newcommand{\siiv}{Si$\;${\small\rm IV}\relax}

\newcommand{\mgii}{Mg$\;${\small\rm II}\relax}

\newcommand{\feii}{Fe$\;${\small\rm II}\relax}

\begin{document}

\title*{Gas Accretion via Lyman Limit Systems}
\author{Nicolas Lehner}
\institute{Nicolas Lehner \at University of Notre Dame, Center for Astrophysics, Department of Physics, 225 Nieuwland Science Hall, Notre Dame, IN 46556, USA \email{nlehner@nd.edu}}
%
%
\maketitle

\abstract{In cosmological simulations, a large fraction of the partial Lyman limit systems (pLLSs; $16 \la \mlnhi < 17.2 $) and LLSs ($17.2 \le \mlnhi < 19 $) probes large-scale flows in and out of galaxies through their circumgalactic medium (CGM). The overall low metallicity of the cold gaseous streams feeding galaxies seen in these simulations is the key to differentiating them from metal rich gas that is either outflowing or being recycled. In recent years, several groups have empirically determined an entirely new wealth of information on the pLLSs and LLSs over a wide range of redshifts. A major focus of the recent research has been to empirically determine the metallicity distribution of the gas probed by pLLSs and LLSs in sizable and representative  samples at both low ($z<1$) and high ($z>2$) redshifts. Here I discuss unambiguous evidence for metal-poor gas at all $z$ probed by the pLLSs and LLSs. At $z<1$, all the pLLSs and LLSs so far studied are located in the CGM of galaxies with projected distances $\la 100$--200 kpc. Regardless of the exact origin of the low-metallicity pLLSs/LLSs, there is a significant mass of cool, dense, low-metallicity gas in the CGM that may be available as fuel for continuing star formation in galaxies over cosmic time. As such, the metal-poor pLLSs and LLSs are currently among the best observational evidence of cold, metal-poor gas accretion onto galaxies. 
}

\section{Introduction}
\label{s-intro}

Among the most pressing problems in galaxy formation is to understand how gas accretion and feedback influence the evolution of galaxies and the intergalactic medium (IGM). Modern cosmological simulations cannot explain the mass-metallicity relationship and color bimodalities observed in galaxies without invoking such large-scale flows \citep[e.g.,][]{tremonti04,keres05}. Stars cannot continue to form in galaxies over billions of years without a replenishment of gas in galaxies from the IGM (e.g., \citealt{maller04,dekel06}), while feedback from star formation and AGN activity can fuel massive outflows from a gas-rich galaxy that may choke off star formation \citep[e.g.,][]{oppenheimer10,oppenheimer12,fumagalli11a,faucher-giguere11,kacprzak08}. The competition between these regulating processes is played out at the interface between the galaxies and the IGM, a region now defined as the circumgalactic medium (CGM).

The large-scale outflows, inflows, recycling in the CGM play therefore critical roles in the evolution of galaxies that require empirical characterization. Outflows appear ubiquitous in the universe at any $z$ \citep[e.g.,][]{pettini01,shapley03,steidel04,steidel10,weiner09,rubin14}. The COS-Halos and COS-Dwarfs galaxy-centric experiments have demonstrated that the low-$z$ CGM is a massive reservoir of galactic metals and baryons with at least as much metals and baryons as in the disk of galaxies (\citealt{tumlinson11a,werk14,peeples14,bordoloi14}, and see also, e.g., \citealt{stocke13,liang14} for other surveys). As discussed at length in this book, gas accretion is crucial for maintaining star formation in the disk of galaxies over billions of years, controlling in part the evolution of the elemental abundances in galaxies, and could be responsible for the mass-metallicity relationship (see, e.g., \citealt{kacprzak16}). However, cold gas accretion that, according to cosmological simulations \citep[e.g.,][]{faucher-giguere11,shen12,faucher-giguere15, fumagalli11a, fumagalli14,vandevoort12a, vandevoort12b,hafen16}, fuels the star formation in galaxies have been extremely difficult to find observationally. According to these simulations,  cold ($T<10^{5.5} $\,K\,$\ll T_{\rm vir}$, e.g., \citealt{vandevoort12a}) gas accretion should not be pristine but already enriched by previous episodes of star formation. However, it should not be either too metal-enriched from outflowing or recycled material from galaxies with intense star formation. Cold streams are also thought to contribute to the population of absorbers with $17 \la \mlnhi <19$, a class of absorbers known as Lyman limit systems (LLSs).  In this chapter, I will argue that the empirically derived properties of the LLSs and partial LLSs (pLLSs) imply that we have found large reservoirs of metal-poor gas around galaxies, but it is still an open question if this gas is actually cold accretion as seen in simulations or is even accreting. 

\begin{figure}[!t]
\includegraphics[scale=0.32]{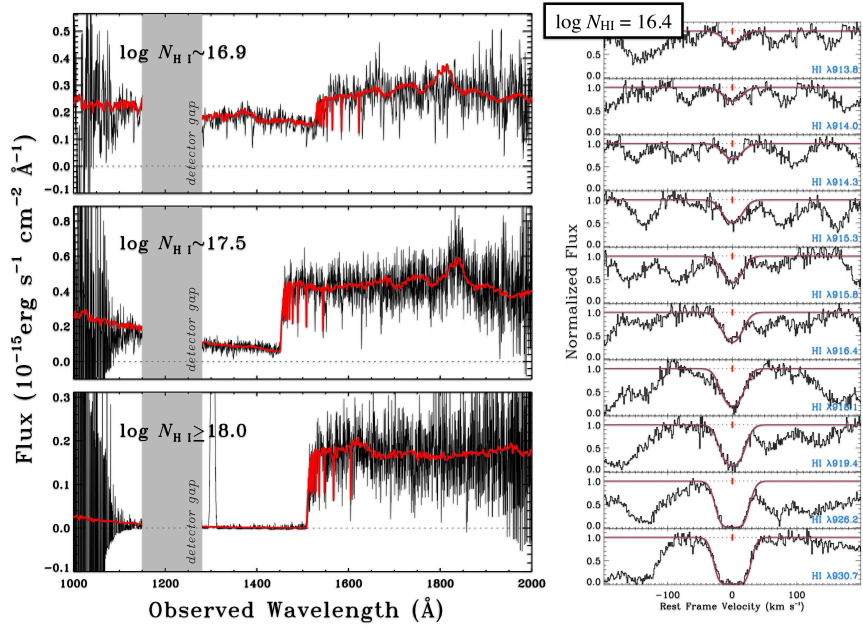}
\caption{To measure \nhi\ in pLLSs/LLSs, two methods are available. 1) {\it left}: we can measure the optical depth at the Lyman limit where  $\tau_{\rm LL} \propto \mnhi$ (these are  low-resolution COS G140L spectra from \citealt{wotta16}; 2) {\it right}: we fit a model to the Lyman series to derive \nhi\ (these are normalized Keck HIRES profiles in the redshift frame of the pLLS at $z_{\rm LLS}\sim 3$ from \citealt{lehner16}).}
\label{f-examples}      
\end{figure}

The pLLSs and LLSs have -- by the definition adopted here -- \hi\ column densities  $16 \la \mlnhi <17.2 $ and $17.2 \le \mlnhi <19 $, respectively, i.e., on the \nhi\ scale of the universe they are located between the \lya\ forest ($\mlnhi <16 $) and the damped \lya\ absorbers ($\mlnhi \ge 20.3 $) and sub-DLAs a.k.a. SLLSs ($19 \le \mlnhi <20.3 $). The \hi\ column densities of the pLLSs and LLSs and the densities implied for the gas they probe ($-4 \la \log n_{\rm H} \la -1$, e.g., \citealt{schaye01,lehner13,fumagalli16}) place them at the interface between the IGM (probed by \lya\ forest absorbers) and galaxies and their immediate surroundings (probed by SLLSs and DLAs). With an optical depth at the Lyman limit $\tau_{\rm LL}\ge 1$, LLSs are optically thick to Lyman continuum radiation, absorbing the UV radiation at $\lambda\le 912$ \AA. Empirically characterizing the UV background has been a scientific driver for studying LLSs as well as their potential connection to galaxies \citep[e.g.,][]{tytler82,sargent89,bahcall93,lanzetta95,omeara07,prochaska10,ribaudo11a}. While pLLSs and LLSs can be readily detected in low resolution spectra and their \hi\ column densities can be estimated directly from the optical depth at the Lyman limit (see Fig.~\ref{f-examples}), these earlier surveys did not have the resolution to detect weak metal lines associated with these absorbers, and hence their properties (metallicity, density, size, mass, etc.) have remained largely unconstrained till more recently. 

The ability to determine the properties of the LLSs and pLLSs is indeed critical for using the pLLSs/LLSs as gas tracers of the IGM/galaxy interface. Among all these properties, the metallicity is key since it can be used as a ``tracer'' of the origins of the gas \citep{ribaudo11,fumagalli11a,lehner13}. The metallicity level of the gas should help differentiating between gas that is accreting from the IGM and the more enriched material produced recently by galaxies at the considered epoch, which could be outflowing or inflowing. If very metal-poor gas is found around a metal-enriched galaxy, it is difficult to find a plausible scenario that would explain how this gas may have come from the galaxy itself. On the other hand, if the circumgalactic gas and the host galaxy have similar metallicities, it is quite natural to invoke large-scale outflows or recycling motions to explain the origin of this gas. 

To determine the metallicity without any bias requires selecting the absorbers based on their \hi\ content alone. Contrary to DLAs (and to a lesser extent SLLSs), many (metal-poor) pLLSs and LLSs at $z<1$ have no detection of \mgii\ $\lambda$$\lambda$2797, 2803 absorption down to equivalent widths $<10$ m\AA; these very weak \mgii\ absorbers are completely missing in even very deep \mgii\ surveys (e.g., \citealt{churchill00}). The \hi\ selection is necessary, but not sufficient since, of course, to be able to estimate the metallicity requires being able to accurately estimate \nhi\ and the column densities of metal lines (which can include non-detections of metals if the signal-to-noise -- S/N -- of the spectra is high enough to place tight constraints on the column densities). It also requires modeling the ionization of the gas since at $\mlnhi \la 18.5$ the gas is predominantly ionized and since only \nhi, not \nhii, can be measured from the observations. 

A first attempt to survey the physical properties of the LLSs was undertaken by \citet{steidel90} with 10 strong LLSs  ($\mlnhi \ga 17$) at $z\sim 3$. He used observations with 35--80 \kms\ spectral resolution, and with this low resolution, the metallicities and other physical properties were only crudely estimated with many limits, but he still was able to show that the total H column of these LLSs is often as large as in DLAs, but mostly in form of \nhii. \citet{prochaska99} showed that with high S/N and high resolution ($\sim$9 \kms) obtained with the 10-m Keck\,I telescope the properties of the LLSs can be estimated much more accurately. The right hand-side of  Fig.~\ref{f-examples} shows that with high resolution and S/N spectra, we can also derive \nhi\ from the analysis of the Lyman series transitions; this is also particularly useful for making sure the metals are observed over a similar velocity range \citep{lehner13,lehner16}.

In this chapter, I will report on recent discoveries within the last $\sim$5--10 years, which in my opinion, have led to several important findings, but have also raised new puzzles. There are two game changers that have occurred in this timeframe, which have enabled transformative progress. As already mentioned high S/N and resolution spectra are important to derive the properties of the pLLSs and LLSs. Equally as important has been the ability to assemble samples of LLSs/pLLSs with more than an handful of absorbers where their properties can be determined. To start to be able to derive some statistically robust trends requires at least a sample size of 25--50. At low redshift, the Cosmic Origins Spectrograph (COS) -- installed in 2009 on the {\it Hubble Space Telescope} (\hst) -- has revolutionized this field thanks to its great leap in sensitivity with spectral resolution $R\sim 17,000$ good enough to accurately estimate \nhi\ and \nmetals. Prior to COS, there was at $z<1$ only a handful of LLSs with a determination of their metallicities (\citealt{lehner09} and references therein). Since the COS installation, the sample has increased to 30--50 \citep{lehner13,wotta16} and will increase to $\ga 100$ in the coming year. At high redshift, dedicated surveys (like, e.g., the HD-LLS survey \citealt{prochaska15}) and the availability of large databases of reduced QSO spectra obtained with large telescopes (e.g., KODIAQ --- \citealt{lehner14,lehner16,omeara15}; XQ-100 --- \citealt{lopez16}) have enabled many new discoveries. 

This chapter is organized as follows. In \S\ref{s-met}, I provide some background on how we determine the metallicities of the pLLSs and LLSs. This process requires making large ionization corrections, but I argue we can do this reasonably well, and well enough to robustly empirically characterize the metallicities of these absorbers. In the next 5 sections (\S\S\ref{s-met-low-z}, \ref{s-high}, \ref{s-mdfcompz1}, \ref{s-pris}, \ref{s-calpha}), I review some of the major findings on the metallicities and relative abundances of the pLLSs and LLSs and their evolution over cosmic time. In \S\ref{s-ovi}, I offer a different but complementary perspective when I present some of the recent results on the surveys of \ovi\ in pLLSs and LLSs and their implications for large-scale outflows and inflows through the CGM of galaxies. In \S\ref{s-acc}, I specifically discuss the findings that may point to cold, metal-poor gas accretion traced by pLLSs and LLSs onto galaxies, and finally in \S\ref{s-summ}, I provide a succinct summary and some future directions. 

\section{Metallicity: methodology and uncertainties }
\label{s-met}
The metallicity is the key parameter to constrain if one wants to determine the properties of the pLLSs and LLSs, assess their possible origins and their associations with some of the structures of the universe. The most robust approach to estimate the metallicity of the pLLSs and LLSs would be to use the \oi/\hi\ ratio since charge exchange reactions with hydrogen ensure that the ionizations of \hi\ and \oi\ are strongly coupled (although in the very diffuse ionized gas probed by these absorbers, some ionization correction may be even necessary when using \oi). However, for absorbers with $\mlnhi \la 17.5$, \oi\ is rarely detected, and the limit that can be placed on $N_{\rm O\,I}$ is generally not sensitive enough to be interesting. So we are left to compare ionized species with \hi, and hence we need to determine \nhii, which requires undertaking ionization modeling of the gas. LLSs and pLLSs are often multiphase, with absorption seen in different ionization stages. However, the low to intermediate ions (e.g., \siii, \siiii, \siiv, \cii, \ciii) and high ions (\ovi) often show distinct kinematics (\citealt{crighton13,fox13,lehner13,lehner14}; and see \S\ref{s-ovi}), with the low ions following more closely the velocity structure of the \hi\ (e.g., \citealt{lehner09,lehner13,crighton13,crighton15, fumagalli16}). The Doppler broadenings of the \hi\ components associated with the pLLSs and LLSs ($b \simeq 15$--30 \kms)  also imply typically that the gas temperature is $ T \la 1$--$4\times 10^4$ K. Hence to estimate the metallicity of the pLLSs and LLSs, one needs to estimate  \nhii\ in the $T\la 1$--$4\times 10^4$ K photoionized gas.\footnote{Except otherwise stated, the metallicity discussed here is the metallicity of the cool gas}. 

\begin{figure}[!t]
\includegraphics[scale=0.7]{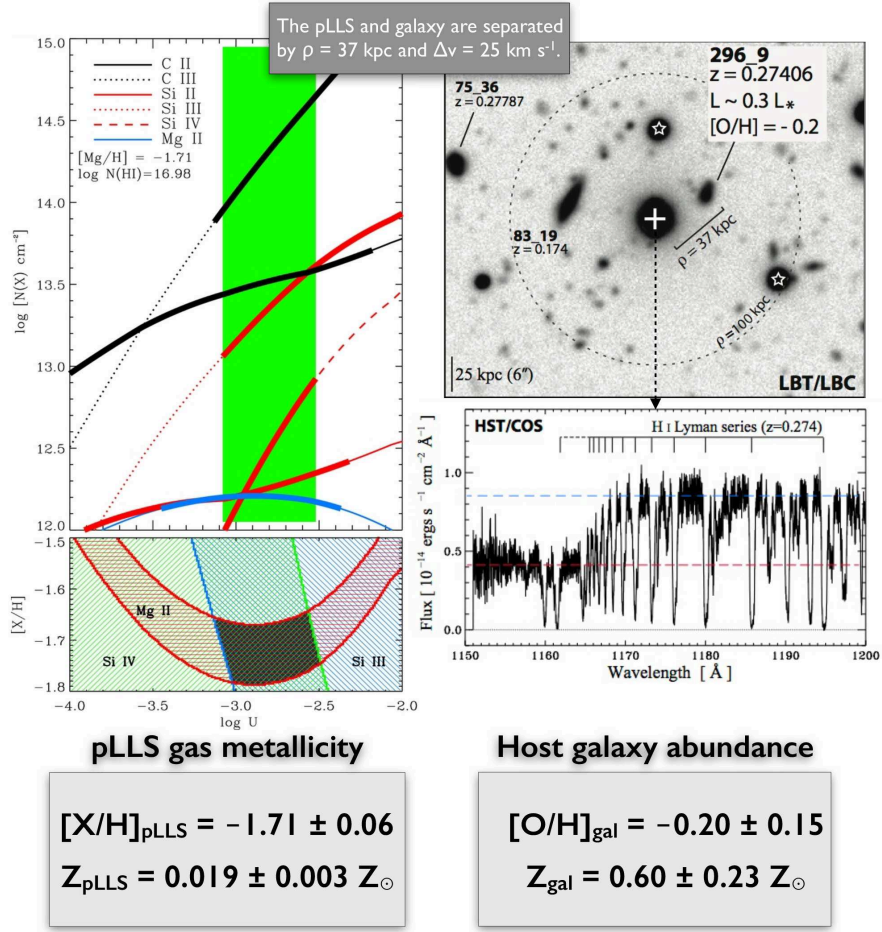}
\caption{One of the first examples of a very low metallicity pLLS found at an impact parameter $\rho = 37$ kpc of a galaxy with the same redshift (results and figures from \citealt{ribaudo11}). The {\it top left panel}\ shows the Cloudy-predicted column densities as a function of ionization parameter  for the given metallicity and estimated \nhi\ (from the COS data shown on the {\it bottom right panel}). The bold portions of the curves show where the model column densities and observations are consistent and the green band shows the range of $U$ for which the models are consistent for all the different ions. The {\it lower left panel}\ shows the allowed $\xh$--$U$ region by the observations. The {\it top right panel} shows a LBT $g$-band image of the QSO field where  the associated galaxy is labeled 296\_9. From a Keck LRIS spectrum, \citet{ribaudo11} were able to estimate an accurate redshift, mass, and abundance of the galaxy. {\it This low metallicity gas clearly did not originate from galaxy 296\_9, and yet it is well within its virial radius.}
}
\label{f-ribaudo}       
\end{figure}

To model the ionization,  it has been customary to use Cloudy \citep{ferland13} where the gas is assumed to have a uniform slab geometry and is photoionized   by the  Haardt-Madau  background radiation field from quasars and galaxies \citep{haardt96,haardt12}. For each absorber, the ionization parameter (the ratio of H ionizing photon density to total hydrogen number density -- $U =n_\gamma/n_{\rm H}$) and the metallicity\footnote{I use throughout the usual square-bracket notation  for the metallicity $\xh \equiv \log N_{\rm X}/N_{\rm H} - \log ({\rm X/H})_{{\odot}} $, where X is a given element.} are varied to search for models that best fit the observational constraints set by the column densities. On the left hand-side  of Fig.~\ref{f-ribaudo}, I show an example of Cloudy simulation for the pLLS  at $z\sim 0.3$ toward PG1630+377 (see \citealt{ribaudo11} for the entire description) where  the  Haardt-Madau  background radiation field from quasars and galaxies (HM05, as implemented within Cloudy) was used. In that case, the upper/lower limits and measurements of the metals constrain well the metallicity. Solar relative abundances are assumed as a starting point for comparing ions of C, Si, O, Mg, etc.,  i.e., a priori the effects of dust or nucleosynthesis on the relative abundances are not included. The quoted metallicity in this chapter is, however, always for the $\alpha$ elements (${\rm X}=$\,Si, O, Mg); I will discuss later that the ${\rm C}/\alpha$ ratio may not always be solar. 

A major drawback to have to correct for ionization is that the characteristics of the photoionized gas depend on the shape and strength of the ionizing spectrum, and there are some systematic uncertainties arising from the ionization modeling. Several independent groups have assessed the systematic errors connected to these ionization models \citep{howk09,lehner13,lehner16,crighton15,fumagalli16,wotta16}. For example, \citet{wotta16} used HM12 \citep{haardt12} to estimate the metallicities of 10 LLSs at $z<1$ from \citet{lehner13} that were initially modeled with HM05. The comparison of the results demonstrates that the harder spectrum of HM12 increases the metallicity  on average by $+0.3$ dex  those derived using HM05 (with no dependence on the initial metallicity of the gas, i.e., a similar shift in metallicity is observed for both high or low metallicity absorbers). For the absorber shown in Fig.~\ref{f-ribaudo}, the metallicity increases by $+0.47$ dex using HM12 instead of HM05, but in the case of HM12, the solution found is not satisfactory because too much \siii\ is predicted by the model ($>+0.2$ dex) while the HM05 background did fit all the observables, including the non-detection of \siii. \citet{fumagalli16} provide a thorough analysis of a large sample of LLSs at $z\sim 2$--3 where they use several ionization models (including dust depletion, proximity to local sources, collisions) and Bayesian techniques to derive the physical properties and metallicities of the LLSs. They find as well that the  metallicity estimates are typically not too sensitive to the assumptions behind the ionization corrections. 

Determining the spectral shape of the ionizing background is an on-going challenge. However, combining all the recent results, there is some consensus that the systematic uncertainties associated with the choice of ionizing background mode are of the order of 0.3--0.5 dex (i.e., a factor 2--3, see references above). The statistical uncertainties for a given model are typically much smaller, especially if the column densities of \hi\ and metals small. Hence, the systematic uncertainty dominates the error budget. This is not precision cosmology, but these uncertainties are comparable to the systematic uncertainties derived using emission line diagnostics for galaxies (e.g., \citealt{berg16}). It is accurate enough to separate confidently a metal-poor absorber ($\xh \la -1$) from an absorber with a near-solar metallicity or to find trends or characterize the metallicity distribution. Furthermore, {\it the same selection criteria and techniques to analyze the characteristics of the pLLSs/LLSs can be used from $z\la 1$ to $z\sim 4$, allowing us to probe the galaxy/IGM interface over 12 billion years.}

\section{The metallicity of the pLLSs and LLSs at $z\la 1$}
\label{s-met-low-z}
Although I cannot yet provide much information about the galaxies associated with the pLLSs/LLSs since this work is still in progress (but see \S\ref{s-acc}), there are great advantages studying the $z\la 1.5$ universe over the higher redshift universe, especially in terms of determining the properties of the associated galaxies. At high redshift, only the brightest, most massive galaxies can be observed, while at $z\la 1$ luminous and very sub-luminous galaxies can be observed. The $z<1.5$ epoch is also  a time where galaxy transformation is occurring with  blue galaxies turning into red galaxies. The morphology, inclination, and geometry of $z<1$ galaxies can also be robustly determined (e.g., \citealt{bordoloi11,kacprzak12b,bouche12}), which will help us providing additional insights on the actual gas origin (see \S\ref{s-acc}). 

In the example shown in Fig.~\ref{f-ribaudo}, I present a pLLS with a very low metallicity $\xh \simeq -1.7$ (2\% solar metallicity) at $z \simeq 0.3$, which is extremely low metallicity at this redshift. This is quite a bit lower than typical of dwarf galaxies in the low redshift universe  (e.g., \citealt{skillman94,kunth00,kniazev03,hirschauer16,berg16}). Fig.~\ref{f-ribaudo} also shows the galaxy environment near the background QSO. A $0.3L*$ galaxy is found at a very similar redshift and with a near solar abundance, a factor $\sim$30 larger than the metallicity of the pLLSs at an impact parameter of $\rho = 37$ kpc. This pLLS is well within the virial radius of this galaxy and yet this gas does not come from this galaxy (at least not from a recent epoch). What is its origin? What is its fate? \citet{ribaudo11} proposed that this pLLS might probe infalling gas onto galaxies, perhaps related to the cold flow accretion observed in cosmological simulations. The physical and chemical properties of this low-metallicity pLLS are comparable to those predicted by cold flow models, including the temperature, metallicity, and \nhi\ as well as host galaxy properties such as velocity offset and mass. This example shows how the metallicity acts as a key discriminator, and since this discovery, one of the goals has been to increase the sample size to determine the prevalence of metal-poor versus metal-rich pLLSs and LLSs at $z<1$. 

At the time of the \citet{ribaudo11} study, there were only about 8 metallicity estimates in LLSs (see, e.g., \citealt{lehner09,cooksey08,zonak04,tripp11}) and about 50\% of those had $\xh \la -1.5$. This is remarkable because as I will show below the sample of pLLSs and LLSs has increased by factors 3.5 and now 7 and yet the same fraction of very metal-poor gas has been found for the combined sample of pLLSs and LLSs. This is also remarkable because these results show the prevalence of low metallicity gas at low redshift where for a long time the canonical value of the metallicity for the IGM was set to $\xh \simeq -1$. The low metallicity has been essentially missed in studies of the \lya\ forest at $z<1$ because for absorbers with $\mnhi \la 15$, the S/N typically achieved in \hst\ UV spectra is not sufficient to detect metals to determine metallicities with $\xh \la -1$. 


\begin{figure}[!t]
\includegraphics[scale=0.53]{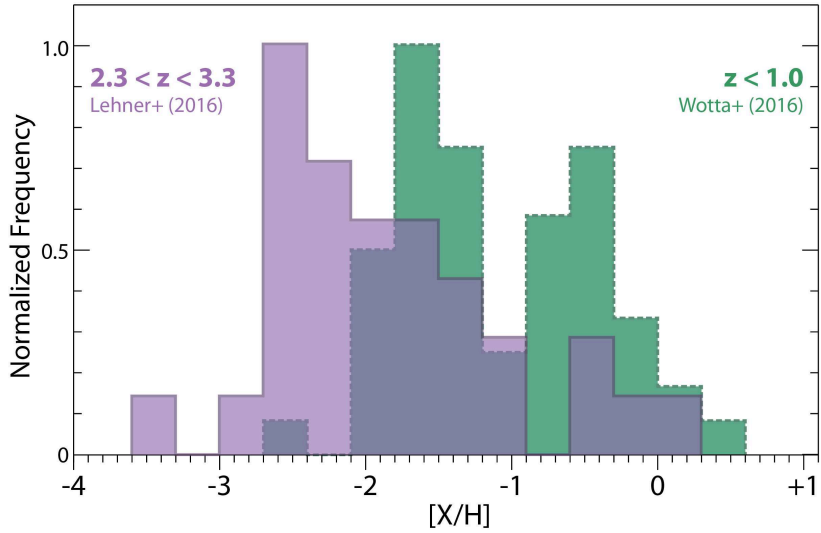}
\caption{Evolution of the metallicity distribution of the \hi\ column density of the pLLSs/LLSs with $16.2 \la \mlnhi \la 18$ from $z<1$ (from \citealt{wotta16}) to  $2.3<z<3.3$ (from \citealt{lehner16}) (figure courtesy of Chris Howk).}
\label{f-comp}       
\end{figure}

One of the main goals of the low-$z$ studies have been to build representative samples of the pLLSs and LLSs. The first sizable sample of pLLSs and LLSs was assembled by \citet{lehner13}. Owing to the original selection of UV-bright QSOs for the high-resolution COS observations, this first sample is heavily biased toward low \nhi\ absorbers with 23 pLLSs and only 5 LLSs. In the second survey, \citet{wotta16} used QSOs that were selected to be optically bright, and hence there is no bias against strong pLLSs or LLSs in the UV spectra of these QSOs, providing a more uniform combined sample in \nhi\ over the pLLS and LLS range. Owing to that selection, the QSOs were obtained with the low resolution mode of COS, leading \citeauthor{wotta16} to develop a new method to derive the metallicity, which used only the comparison \mgii\ and \hi. They demonstrate that with a prior knowledge of the $U$ distribution (determined from the analysis of \citealt{lehner13}), the metallicities of the pLLSs and LLSs at $z\la 1$ can be accurately estimated and the two samples from \citet{lehner13} and \citet{wotta16} can be combined.  I refer the reader to the original papers for more details. Here I use the combined sample from the two surveys  (which also include the few LLSs with $\mlnhi \ge 17.7$ which were not included in \citealt{wotta16}). 

\begin{figure}[!t]
\includegraphics[scale=0.35]{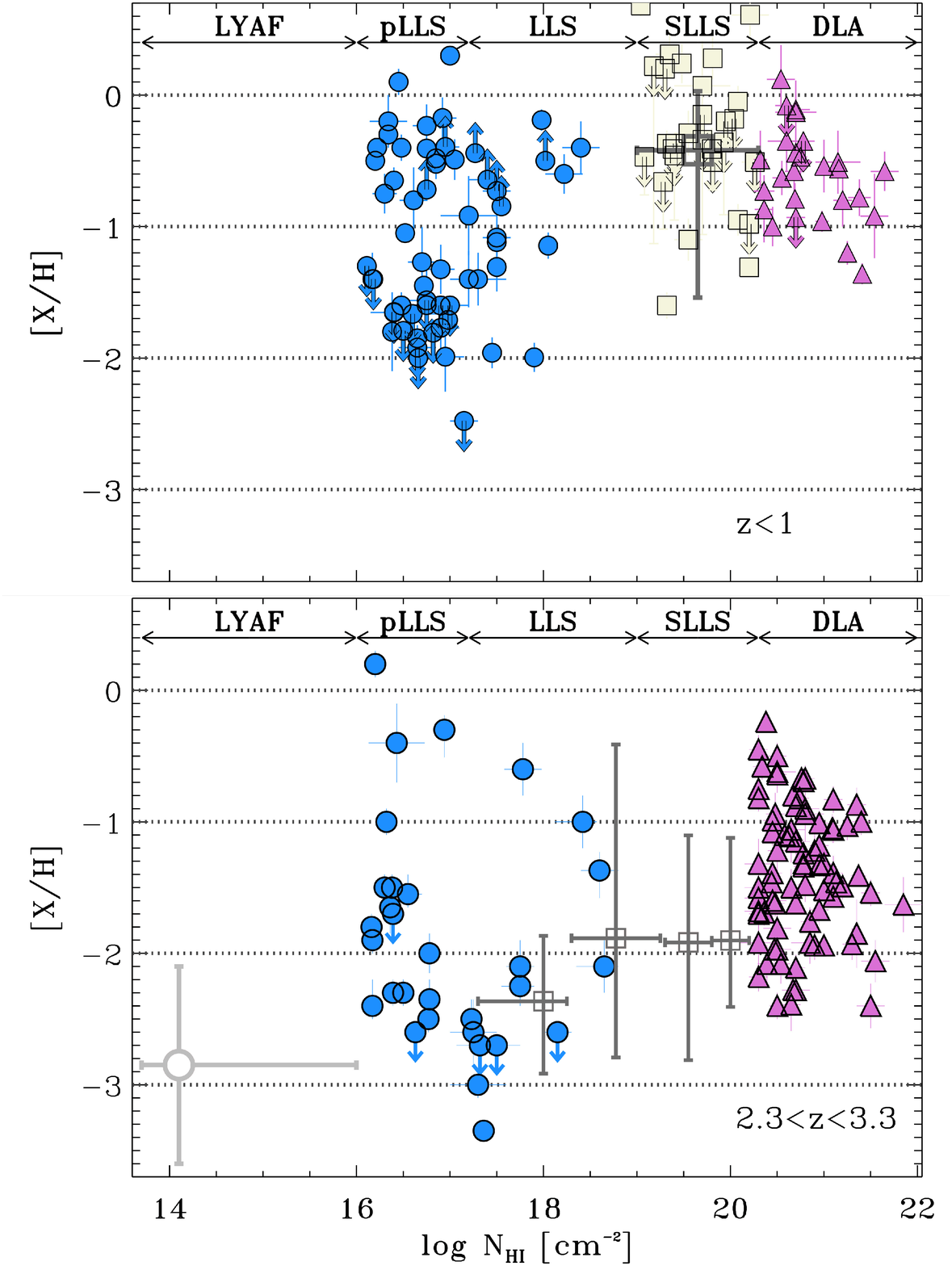}
\caption{Metallicity as a function of the \hi\ column density for absorbers  at $z<1$ ({\it top}, results from \citealt{lehner13,wotta16} and references therein) to  $2.3<z<3.3$ ({\it bottom}, results from \citealt{lehner16} and references therein, and see also the main text). Note that the metallicity-scale in both panels is the same for direct comparison. 
}
\label{f-metvsnh1}       
\end{figure}

Figs.~\ref{f-comp} and \ref{f-metvsnh1} present two different ways to display the metallicity distribution of the pLLSs and LLSs. The green histogram in Fig.~\ref{f-comp} shows the metallicity distribution of the combined sample of pLLSs and LLSs at $z<1$, with two prominent peaks at $\langle \xh \rangle \simeq -1.9$ and $\langle \xh \rangle \simeq -0.3$ (see \citealt{wotta16} and \citealt{lehner13} for details; the low-metallicity peak has a mean that appears low relative to its distribution peak because it was estimated using the survival analysis to take into account of the 13 upper limits that are displayed in Fig.~\ref{f-metvsnh1}). A Gaussian mixture modeling (GMM) and a Dip test on the metallicity distribution of the pLLSs and LLSs rejects a unimodal distribution at high significance levels (99.6\% and 82.4\%, respectively). As noted \citet{muratov10}, the Dip test appears less powerful than GMM  based on the significance levels, but it has the benefit of being insensitive to the assumption of Gaussianity and is therefore a true test of modality. 

Fig.~\ref{f-metvsnh1} shows the metallicity of the absorbers as a function of \nhi. This type of figure allows us to make a direct comparison with other \nhi\ absorbers as well as displaying the unbinned metallicity distribution of the absorbers. While the connection between \nhi\ and impact parameter to the galaxies is crude, examining the metallicity for absorbers covering a broad range of \hi\ column densities provides a measure of how the origins of the gas are changing with environment (impact parameter and density) within galaxy halos. I will discuss this evolution with \nhi\ further in \S\ref{s-mdfcompz1} and focus here solely on the pLLSs and LLSs.

Starting with pLLSs, there is a lack of absorbers in this figure near $\xh \sim -1 $ ($-1.2 \la \xh \la -0.9$). For the pLLS sample, the Dip test reject an unimodal distribution at the significance level of 95.5\% ($>99.9\%$ according to the GMM), i.e., it is more significant than for the combined pLLS and LLS sample.\footnote{In the Dip test and GMM, upper and lower limits are treated as values. If one decreases/increases the upper/lower limits by a very large factor $\ge 5$ (0.7 dex), the unimodal distribution is still rejected at the $\ge 80\%$ and $\ge 95\%$ significance level for the Dip test and GMM, respectively. Therefore, the metallicity distribution of the pLLSs is very likely at least bimodal (there could be another population of pLLSs at $\xh \la -2$ if the actual values of the several of the upper limits are indeed much lower than the current limits set by the observations).} Fig.~\ref{f-metvsnh1} also reveals that while bimodal nature of the metallicity distribution is now pretty well established for the pLLSs\footnote{Note as well that the upper and lower limits push the two branches apart, not together}, it is less so for the LLSs. In fact there is a strong hint that the metallicity distribution may transition to a broad unimodal distribution in the LLS regime \citep{wotta16}: 1) there are more LLSs near the $-1$ dex dip, and 2) there are significantly less LLSs with $\xh \le -1.4$ than pLLSs (5--19\% compared to 35--51\% for the pLLSs; 68\% confidence level ). An ongoing survey will double the current sample of the LLSs using the COS archive and show if indeed the metallicity distribution changes in the LLS regime (C. Wotta et al. 2017, in preparation). While there is lack of LLSs with $\xh \le -1.4$, the fraction of LLSs with $\xh \le -1$ is 39--61\% (68\% CL) is still quite important. 

\section{Metallicity distribution at high $z$ and redshift evolution}
\label{s-high}

The great aspect with QSO absorbers is that the same diagnostics and techniques are used at both high and low $z$, so the evolution of the metallicity and physical conditions of the absorbers can be studied straightforwardly over cosmic time. The cosmic evolution of the DLAs  \citep[e.g.,][]{prochaska03,rafelski12,battisti12,jorgenson13}  and SLLSs  (e.g., \citealt{som13,som15,fumagalli16,quiret16}) have been studied for several years. Thanks to large archives of good quality (in particular high resolution) spectra of both low and high $z$ QSOs, we can now study the cosmic evolution of the \hi-selected pLLSs and LLSs. The same selection criteria and analyses of the pLLSs and LLSs are applied at $z<1$ and $2.3<z<3.3$ by \citet{lehner13} \citep[and also][]{wotta16} and \citet{lehner16}, respectively, which allow for a direct comparison between the two samples as displayed in Figs.~\ref{f-comp} and \ref{f-metvsnh1}. They are \hi-selected to have \hi\ column densities between $16 \la \mlnhi <19$. The \hi\ column density for each absorber can be estimated reasonably accurately (within $\sim$0.3 dex, and often better than 0.05--0.10 dex). There is enough information from the metal lines to derive sensitively both high and low metallicities. And finally, the reported metallicities are from  $\alpha$-elements, and i.e., ${\rm X} = \alpha$ where $\alpha$ is Si and/or O at high $z$ and Si, Mg, S, and/or O at low $z$. 

The reader should, however, keep in mind that the overdensities of the structures of the universe change as a function of $z$, and therefore absorbers at some given \nhi\ at high and low redshifts are not necessarily physically analogous \citep[e.g.,][]{schaye01,dave99}. The change in the overdensity $\delta_{\rm H} \equiv (n_{\rm H} - \bar{n}_{\rm H})/\bar{n}_{\rm H}$ is a factor $\sim$8  between the mean redshifts of the low and high $z$ samples \citep{lehner16}. For the \lya\ forest absorbers, SLLSs, and DLAs, the redshift evolution of the density does not change the fact that  \lya\ forest absorbers still trace very diffuse gas ($\delta_{\rm H}\ll 100$) and SLLSs/DLAs trace virialized structures ($\delta_{\rm H}\gg 100$) at any $z$. However, for the LLSs and especially the pLLSs, while at $z<1$ they probe gas well within the CGM of galaxies (see \S\ref{s-acc}), at $z\sim 2.8$, $\delta_{\rm H}$ can be $\la 100$, and hence some pLLSs could probe more diffuse ionized gas at $z>2$. In terms of high redshift galaxy observations, KBSS shows that only half of the absorbers with $\mlnhi >15.5$ are found in the CGM of ({\it massive}) galaxies at $z\sim 2$; the other half may probe more diffuse gas or the CGM of dwarf galaxies \citep{rudie12}. Observations with MUSE also found no bright, star-forming galaxy in the vicinity of two very metal-poor LLSs \citep{fumagalli16a}. However, I emphasize these observations targeted only very metal-poor LLSs, which are just a subset of the LLS population (see below). With the current knowledge, while LLSs and pLLSs are by definition at the interface between the denser and more diffuse gas at any $z$, high $z$ pLLSs and LLSs may not always trace necessarily the dense CGM of galaxies as their counterparts at $z<1$. 

In Figure~\ref{f-comp}, I show the metallicity distribution for the 31 \hi-selected pLLSs and LLSs at $2.3 < z < 3.3$ from the KODIAQ Z pilot study \citep{lehner16}. Visually and statically, the metallicity distribution at high $z$ is quite different from the low $z$ metallicity distribution: not clearly bimodal and the peak of the distribution has shifted to lower metallicity. \citet{lehner16} estimate for the pLLSs {\it and}\ LLSs that  $\langle {\rm [X/H]}\rangle = -2.00 \pm 0.17$ (where the quoted error is the Kaplan-Meier (KM) error on the mean value obtained from the survival analysis).  There is little overlap at $\xh \le -2$ between the high and low redshift samples, although there are several upper limits around $-2$, $-1.7$ dex at $z<1$ (see Fig.~\ref{f-metvsnh1}). However, at $\xh \ge -2$, there is a large overlap. Quite remarkably there are also several pLLSs/LLSs (with a fraction $\sim$\,4--20\%, 68\% CL) with $-0.5\la \xh \la +0.2 $ at $z>2$.  There is no evidence of a strong dip anywhere in the metallicity distribution as observed at low redshift (L13, W16)
and there is a prominent peak near the mean. The Dip test on the high $z$ metallicity distribution shows that the significance level with which a unimodal distribution can be rejected is only 26\%. 

In Fig.~\ref{f-comp}, I also show the result from the HD-LLS survey with the square gray data points from \citet{fumagalli16} over the same redshift interval than probed by the KODIAQ Z sample shown in this figure. The HD-LLS survey targets LLSs and SLLSs with $\mlnhi > 17.2$ at $z\sim 2.5$--3.5  \citep{prochaska15}, i.e, over a similar redshift interval. There are only 9 LLSs with $\mlnhi \sim 17.5$, while all the others have $\mlnhi \ga 18$. Yet it is striking how the averaged values for overlapping \nhi\ agree remarkably well (see Fig.~\ref{f-comp}; I emphasize that point because even though there are similar authors on both the HD-LLS and KODIAQ Z teams, the results were independently derived). 

Finally, as I discuss above the pLLSs and LLSs at low $z$ may not sample the same structure as their high redshift analogs. However, the LLSs at $2.3<z<3.3$ could evolve into the low $z$ pLLSs. From the results displayed in Fig.~\ref{f-comp}, it is evident that the metallicity distribution of the LLSs at $2.3<z<3.3$ is consistent with a unimodal distribution, significantly different from the bimodal metallicity distribution of the pLLSs at $z\la 1$. Therefore, even considering the redshift evolution of the cosmic structures, there is a significant evolution of the metallicity distribution of the LLSs with $z$.

\section{Metallicities as a function of \nhi\ over cosmic time}
\label{s-mdfcompz1}

As alluded to before, plotting the metallicity as a function of \nhi\ allows for a direct comparison on how the metallicity distributions change with \nhi\ from the most diffuse regions to densest regions of the universe traced by \hi, and, with information at different $z$, we can determine how it evolves over cosmic time.  At $z<1$, the S/N of the COS QSO spectra does not allow us to probe sensitively metallicity below $\xh \la -1$ for the \lya\ forest, i.e., absorbers with $\mlnhi \la 15$. However, at $z>2$ high S/N QSO spectra can be obtained so the metallicities of the \lya\ forest can be determined with much less bias  \citep[e.g.,][]{ellison00,schaye03,aguirre04,simcoe04}. In the right-hand side of Fig.~\ref{f-metvsnh1}, I show the mean and standard deviation from  \citet{simcoe04} who determined in the spectra of 7 QSOs the metallicity using \ovi\ and \civ\ for absorbers with $13.6 \la \mlnhi \la 16$ (most between  $13.6 \la \mlnhi \la 14.4$ which is highlighted by the asymmetric error on the horizontal axis) at $z\sim 2.5$. In that figure, the gray squares represent the mean metallicity of the strong LLSs and SLLSs from the HD-LLS survey \citep{fumagalli16}, where the horizontal bar indicates the range of the \nhi\ bin and the vertical error bar represents the 25th/75th percentile of the composite posterior PDF. For the DLAs, I use the measurements and compilation from \citet{rafelski12}. For the SLLSs and DLAs, see also the compilation made by \citet{lehner13} and \citet{quiret16} and references therein (for the \citeauthor{quiret16} sample, please note that there has been no attempt to remove or flag absorbers that were selected with some a priori, e.g., very weak metal features in SDSS spectra). 

As displayed in Fig.~\ref{f-metvsnh1} and reviewed in length by \citet{lehner13} and \citet{wotta16}, it is clear the metallicities vary strongly with \nhi\ at any $z$. Considering the strong \hi\ absorbers with $\mlnhi \ge 16$, it is readily apparent that the prevalence of low metallicity gas is only observed for the PLLSs and LLSs.  We can quantify this further by defining the ``very metal-poor'' (VMP) absorbers as those with metallicities 2$\sigma$ below the mean metallicity of the DLAs in any given redshift interval. At $z<1$, that threshold is $\xh_{\rm VMP} \le -1.4$ while at $2.3<z<3.3$, it is $\xh_{\rm VMP} \le -2.4$. By definition, VMP gas is quasi-absent for DLAs ($<3\%$--10\%, 68\% CL, which applies for all the percentage given here), but it is also small for the SLLSs ($<10\%$ at $z<1$; $\sim 20\%$ at $2.3<z<3.3$). On the other hand, the fraction of VMP pLLSs+LLSs at any $z$ is much larger (20\%--45\%) and with the current statistics relatively invariant over cosmic time from $z<1$ to $z\sim 2$--3.  There are some hints that the pLLSs and LLSs may have different metallicity distributions at both low and high $z$, but this will require increasing the sample sizes to robustly determine these distributions across the \nhi\ range probed by the pLLSs and LLSs. 

Therefore at any $z$, VMP gas is only observed in pLLSs and LLSs and the \lya\ forest (presumably that is the case too at $z<1$ for the \lya\ forest). The VMP gas must have a different origin than the typical DLAs. It demonstrates there are very large reservoirs of VMP gas around $z<1$ galaxies and in the interface between the IGM and galaxies at $z>2$.   At $3.2 \le z\le 4.4$ with a smaller sample probing very strong LLSs ($17.8 \la \mlnhi \la 19.5$)  and an indirect method, \citet{cooper15} found  28\%--40\% of the LLS population could trace VMP gas.\footnote{\citeauthor{cooper15} (and see also \citealt{glidden16}) targeted a sample of 17 high \nhi\ LLS  (typically $\mlnhi \sim 17.5$) at $3.2\la z \la 4.4$ , but selected them on the absence of metal absorption in SDSS spectra, i.e., they single out them to be a priori low-metallicity LLSs.} The metallicities of the VMP pLLSs/LLSs increase with decreasing redshift, but their fractions appear to remain about the same over 12 billion years.

On the other hand, about half of the sample of pLLSs and LLSs at both high and low $z$ has metallicities that overlap with the DLA and SLLS populations, implying that metals are found in large quantities in large and small overdensities of the universe and near and far from galaxies. Since pLLSs and LLSs are typically probing gas beyond the luminous boundaries of galaxies, the origins of these pLLSs and LLSs must have been originally outflows or ejection of metals via interactions of galaxies. Gas recycling (i.e., infalling gas previously ejected from a galaxy) is also another possibility for these metal rich pLLSs and LLSs. 

\section{Pristine LLSs}
\label{s-pris}
Although cosmological simulations predict that LLSs probing cold flow accretions should have been enriched to some levels \citep[e.g.,][]{fumagalli11b,hafen16}, with the observations we can estimate the importance of pristine pLLSs and LLSs. My definition of pristine gas is no detection of metals and an upper limit on the metallicity $\xh < -3.5$ dex. At $z<1$, there is currently no pLLS or LLS that fits this definition. In the \citet{lehner13} sample, several pLLSs and LLSs have only upper limits on the metallicity, but the most sensitive one is $\xh <-2$. For this absorber, there is detection of \ciii\ and \oiii. In the \citet{wotta16} sample, there are also several pLLSs and LLSs with upper limits on $\xh$. However, for that study only \mgii\ was used to determine the metallicity. \mgii\ is certainly one of the most critical ions to determine the metallicity at $z<1$ for the pLLSs and LLSs \citep{lehner13}, but its non-detection does not imply that the gas is pristine since in all known cases so far there has been always absorption of \ciii, \siiii, and/or \oiii\ when these ions are available. And, indeed, in the lowest metallicity pLLS of the \citet{wotta16} sample ($\xh <-2.5$), there is likely a detection of \ciii\ absorption found in the NUV spectra with follow-up \hst/COS G185M observations (J.C. Howk et al. 2016, in prep.). Future efforts to improve these limits and probing a larger sample of LLSs at $z<1$ will be critical to determine the lowest levels of metal enrichment in the low redshift universe. 

At $z\sim 3$, there are currently only two examples of LLSs with no evidence of metal absorption down to a limit ${\rm [X/H]}< -4.2$ and $< -3.8$ \citep{fumagalli11b}. \citet{lehner16} detected several pLLSs and LLSs with no metals that might be reminiscent of these pristine  LLSs, but, unfortunately, \siiii\ is contaminated for each of these cases, and no stringent limit on the metallicity can be set. To better understand the level of mixing of metals in the early universe,  we will need a larger sample to reliably determine the frequency of pristine gas at $2<z<4.5$ in the interface regions between galaxies and the \lya\ forest absorbers.  From the current sample, the fraction of pLLSs/LLSs with ${\rm [X/H]< -3.5}$ with {\it no}\ metal detected is $<7.8\%$ at 90\% CL. 

\section{C/$\alpha$ in pLLSs and LLSs over cosmic time}
\label{s-calpha}

 Although we have limited information on the relative abundances, we have some constraints on the C/$\alpha$ ratio, which is a good indicator of the nucleosynthesis history (since in low density, diffuse gas, carbon is not expected to be strongly depleted into dust grains), and hence this ratio provides additional information regarding the origin of the gas. For the pLLSs and LLSs, this ratio was principally derived from the photoionization models, but \citet{lehner16} also report direct measurements using the different ionization stages of C and Si [for example at high $z$, ${\rm C}/\alpha= (N_{\rm CII} + N_{\rm CIII} +N_{\rm CIV})/(N_{\rm SiII} + N_{\rm SiIII} +N_{\rm SiIV})$], and quite remarkably the direct and modeling methods provide consistent results.

\begin{figure}[!t]
\includegraphics[scale=0.5]{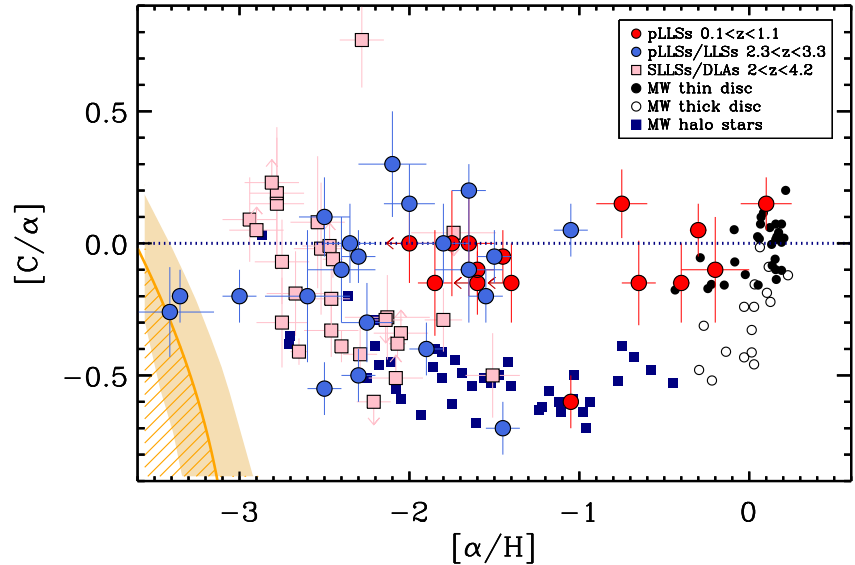}
\caption{Evolution of  [C/$\alpha$] as a function of the metallicity $[{\rm \alpha/H}]$ for various types of absorbers at different $z$ and stars indicated in the legend (from \citealt{lehner16}). Any absorbers in the  hatched orange region may have been polluted by Pop III stars. 
}
\label{f-calpha}       
\end{figure}

For the pLLSs and LLSs, at high $z$, $\alpha$ is mostly Si, but at low redshift it can also be O, Mg, and/or S depending on the system and redshift. In Fig.~\ref{f-calpha}, I show [C/$\alpha$] vs. [$\alpha$/H] for the pLLSs and LLSs from both the high- and low-redshift samples (from \citealt{lehner16}). For comparison, the results for high redshift DLAs and SLLSs and Milky Way stars are shown. For the DLAs and SLLSs, the results are from \citet{pettini08}, \citet{penprase10}, and \citet{cooke11a} (and references therein and see also \citealt{becker12} for $z\ga 5$ measurements). For the Galaxy thin and thick stars, the results from \citet{bensby06} are used, and for the halo stars, \citet{fabbian09} and \citet{akerman04}. For the stars, $\alpha$ is O, while for the DLAs and SLLSs, $\alpha$ is O or Si (changing O to Si or vice-versa for the DLAs would have little effect on the distribution of these data). 

The overall trend observed in Fig.~\ref{f-calpha} in the stellar and SLLS/DLA samples can be separated in roughly two regions: {\it region 1} covering $-3\la [{\rm \alpha/H}] \la -1$, where [C/$\alpha$] decreases with increasing metallicity from super-solar to about $-0.7$ dex;  and {\it region 2} over $-1\la [{\rm \alpha/H}] \la +0.2$, where [C/$\alpha$] increases with increasing metallicity from about $-0.6$ dex to super-solar. The behavior in region 2 has been well known for some time and is thought to occur as a result of the delayed release of carbon from low- and intermediate-mass stars combined with a strong metallicity dependence of the yields of carbon by massive stars with mass-loss \citep[e.g.,][]{akerman04,fabbian09}. The increase of C/$\alpha$ to lower metallicity at $ [{\rm \alpha/H}] \la -1$ has now been confirmed independently in both stellar atmospheres and SLLSs/DLAs. The exact reason for this trend is not yet well understood. One possible interpretation for the high values of C/$\alpha$ could  be the leftovers from the enhanced production of C (relative to $\alpha$-elements, and in particular O) in Population III (Pop III) stars. As shown by \citet{frebel07} and \citet{bromm03}, the gas progenitor of Pop III stars must have high C abundance to efficiently cool the gas in order to actually form stars and drive the transition from Pop III to Pop II stars (see also \citealt{cooke11a} for more discussion). This condition is shown in Fig.~\ref{f-calpha} (hatched orange region), which is defined as the ``transition discriminant" criterion. No Pop II stars should be found in that zone, but any gas in this region will likely have been polluted by Pop III stars; as we discuss below and can be seen in Fig.~\ref{f-calpha}, only two LLSs are found in that ``forbidden'' zone.

For the pLLSs and LLSs, things are quite different. First of all it is worth to note that in the regions of overlapping metallicities, there is no evidence of a difference between the low and high redshift samples, and hence they can be treated as one. About half the sample of the pLLSs and LLSs follows a similar distribution to that observed for the DLAs and stars over the entire range of metallicities. For these, their chemical enrichment history (at least of C and $\alpha$-elements) appears to be similar to that of the Galactic stars and the bulk of the SLLSs/DLAs. However, the other half --- mostly clustered at $-2.2\la [{\rm \alpha/H}] \la +0.2$ --- does not follow the trend observed in stars or DLAs.  These gas clouds are carbon-enhanced by a factor $\ga 2$--5 ($\ga 0.3$--$0.7$ dex) compared to stars or most DLAs with similar  $[{\rm \alpha/H}]$. The enhanced  C/$\alpha$ ratio in the metallicity range $-2 \la {\rm [X/H]\la -0.5}$ implies that this gas must have been polluted by preferential ejection of C from low metallicity galaxies. A recent study in fact shows that at least some local metal-poor dwarf galaxies have also enhanced C/$\alpha$ over similar metallicities \citep{berg16}. While their  C/$\alpha$ ratios are not as high as observed for the pLLSs and LLSs and their sample is small (12 galaxies), the absence of a clear trend  between C/$\alpha$ and $\alpha$/H  is similar to that observed in pLLSs and LLSs.

As noted by \citet{crighton16}, extremely metal-poor LLSs ( ${\rm [X/H]}\sim -3.5$ at $z\sim 3$) with detected metal absorption also provide another path to study the Pop III/Pop II metal-enrichment transition. There are now two LLSs at $z\sim 3.4$ with expected [C/$\alpha$] and ${\rm [\alpha/H]}$ that are consistent with gas polluted from Pop III stars (\citealt{crighton16,lehner16}; see Fig.~\ref{f-calpha}).  The use of both the low metallicity and C$/\alpha$ ratio provides a strong method to find metal-pollution at the transition from Pop III to Pop II star formation  \citep[see][]{cooke11a}. 

\section{Lyman Limit systems and \ovi}
\label{s-ovi}

Although I have focused throughout on the metallicity of the cool gas of the pLLSs and LLSs, the surveys described above have also revealed that  \ovi\ absorption with overlapping velocities with \hi\ is found with a high frequency at any $z$. At $z<1$, among the 23 pLLSs/LLSs with \ovi\ coverage, only 3 have no \ovi\ absorption below the lowest \novi\ of detected \ovi\ absorption (there are an additional 3, but the limits were not sensitive). Hence the detection rate of \ovi\ absorption associated with pLLSs/LLSs at $z<1$ is about 70\%--80\% \citep{fox13}.  At $2.3<z<3.6$, the KODIAQ survey found that about 70\% of the LLSs also show \ovi\ absorption \citep{lehner14}, but this is still based on small-number statistics (there are only 7 LLSs in the KODIAQ I survey, the other absorbers being SLLSs and DLAs, but see also, e.g., \citealt{kirkman97,kirkman99}).

The \ovi\ absorption is particularly important because owing to its high ionization energies (114--138 eV), it is difficult to photoionize by the UV background alone except if the gas is extremely low density ($n_{\rm H}\la 10^{-3.5}$ \cmmm, e.g., \citealt{simcoe02}). When \ovi\ is detected, these pLLSs and LLSs have typically multiple gas-phases as evidenced by the presence of low ions (e.g., \cii, \siii, \siiii) and \ovi\ (or other high ions) that cannot be explained by a single photoionization model (e.g., \citealt{lehner09,lehner13,crighton13}). Furthermore and independent from any ionization modeling, the high-ion and low-ion gas-phases of pLLSs/LLSs are often distinct based on the kinematic comparison of their velocity profiles, which indicate that most of the pLLSs/LLSs have kinematic sub-structure with velocity centroid offsets between the strongest \hi\ (or low ions) and \ovi\ absorption and/or multiple components seen only in the low or high ions \citep{fox13,crighton13,lehner14}. The high S/N and resolution of the KODIAQ spectra also demonstrate that the \ovi\ profiles have typically a larger velocity breadth with the bulk of their absorption often offset from the low ions by several tens or hundreds of kilometers per second and are generally smoother (i.e., with many less velocity components) \citep{lehner14}. These properties contrast remarkably with those of the  \lya\ forest ($\mlnhi \la 14.5$) where  for  a majority of systems, \ovi\ and \hi\ coexist in the same gas with similar (and generally much simpler) kinematics \citep[e.g.,][]{tripp08,muzahid12,savage14}.  

\begin{figure}[!t]
\includegraphics[scale=0.32]{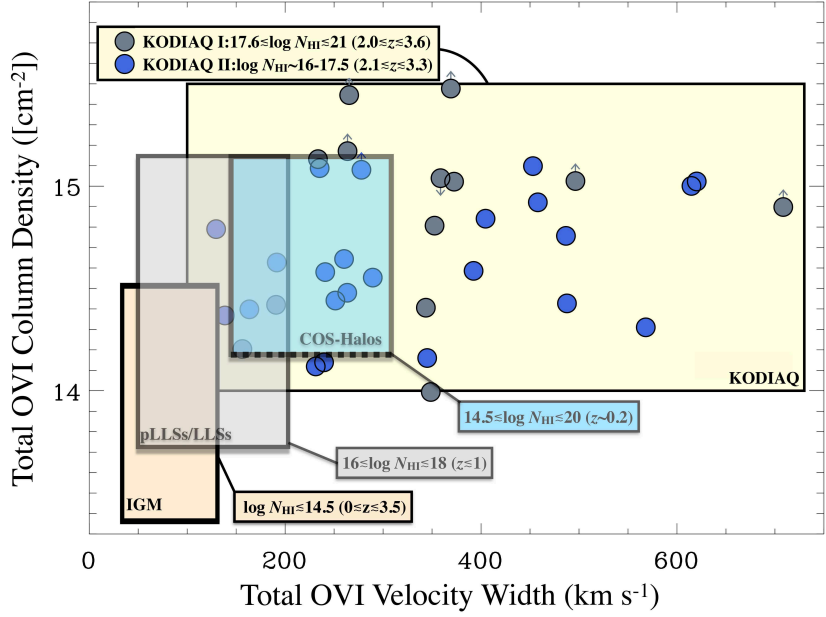}
\caption{Total column density of \ovi\ ($\mlnovi $) versus the total width ($\Delta v_{\rm OVI}$) of the \ovi\ absorption found in different surveys. The rectangles show where most of the data lie for each survey. For the KODIAQ and the $z<1$ pLLS/LLS surveys (sample sizes $m \simeq 36$, 23, respectively), the absorbers were \hi-selected to satisfy the range of \nhi\ values listed in the figure. For COS-Halos ($m\simeq 40$), absorbers were associated with pre-selected $\sim L*$ galaxies so that they are at impact parameters $\rho\la 160$ kpc (note that dotted lines show about the sensitivity level of the COS-Halos survey in \novi; for \hi-selected surveys, the sensitivity is typically much better than the lowest values reported in this figure, around $\mlnovi \la 13.2$--13.5). For the IGM \lya\ forest surveys ($m\ga 50$-100), absorbers are typically \ovi-selected found to be associated with low \nhi\ absorbers.  }
\label{f-ovi}       
\end{figure}

Although \ovi\ is traditionally associated with outflows (see below), in cosmological simulations some of the gas accretion is expected to occur in the form of \ovi-bearing gas (see, e.g., \citealt{shen13,ford14,ford16}). Specifically, in the Eris2 simulations of a massive star-forming galaxy at $z\sim 3$, outflows and inflows have similar covering factors within $\sim 1 R_{\rm vir}$ for \ovi\ with $13.5 \le \mlnovi < 14.4$, while only outflows have a large covering factor for absorbers with  $\mlnovi \ga 14.4$ at any impact parameters  within 1--$2R_{\rm vir}$ \citep{shen13}. Further insights can be gained by directly comparing different surveys, and an easy and informative way to do this is by considering the total integrated properties of the \ovi\ absorption, i.e., the total column density (\novi) and the total full-width ($\Delta v_{\rm OVI}$). In Fig.~\ref{f-ovi}, I show a summary of the current status of $\mlnovi$ vs. $\Delta v_{\rm OVI}$ for the IGM at both low and high redshift \citep[e.g.,][]{simcoe02,muzahid12,tripp08,savage14}, the COS-Halo survey \citep{tumlinson11a,werk13,werk16}, the KODIAQ survey of high-$z$ LLSs (\citealt{lehner14,burns14}; N. Lehner et al. 2017, in prep.), and the $z<1$ pLLS/LLS survey \citep{fox13,lehner13}. The rectangles in this figure show where most of the data lie for a given survey. 

For the KODIAQ survey (and the $z<1$ pLLS/LLS survey), all the absorbers are pre-selected based on their \hi\ content so that they satisfy some \nhi\ threshold (KODIAQ I, $\mlnhi \ga 17.5$; KODIAQ II, $\mlnhi \ga 16$ with typically most of the absorbers with $16 \la \mlnhi \la 17$, i.e., the \nhi\ distribution of the KODIAQ survey II is more similar to that of the $z<1$ survey).\footnote{Note also that for KODIAQ I, an additional criterion is that the metallicity of the cool gas could be determined, which is not the case for KODIAQ II.} Since the KODIAQ II results are not yet published, I show their actual distribution. When \ovi\ is detected, most of the KODIAQ data are such that  $14.2 \la \mlnovi \la 15.5$ and $150 \la \Delta v_{\rm OVI} \la 500$ \kms. More than half the KODIAQ I+II sample has $\mlnovi \ga 14.4$ and $\Delta v_{\rm OVI} \ga 300$ \kms, i.e., a majority fits in this category of strong and broad \ovi\ absorption that is associated with large-scale outflows in cosmological simulations. As noted by \citet{lehner14},  the breadth and strength of the \ovi\ absorption in optically thick \hi\ absorbers at $z\sim 2$--3.5 are striking and quite similar to those observed in starburst galaxies in low redshift universe \citep[see, e.g.,][]{grimes09,tripp11,muzahid15}, strongly suggesting that indeed the origin of the bulk of the strong \ovi\ absorption associated with pLLSs and LLSs is hot cooling gas of large-scale outflows from high-redshift galaxies at an epoch where galaxies are forming stars at very high rates.  Fig.~\ref{f-ovi} also shows that the KODIAQ and COS-Halos\footnote{As a reminder, COS-Halos is a dedicated absorption-line survey of $\sim L*$ galaxies at $z\sim 0.2$ where galaxies have various of SFRs and morphologies. The background quasars probed their CGM at impact parameters $ 10\la \rho\la 160$ kpc \citep{tumlinson11a,tumlinson13,werk12,werk13,werk14,werk16}. The \ovi\ absorption is only observed in the CGM of star-forming galaxies with a sensitivity limit $\mlnovi \ga 14.2$.} surveys overlap, but are remarkably outside the boundaries set by the \lya\ forest.\footnote{The \ovi\ absorption in the IGM has been searched from blind \ovi\ surveys. The properties of the \ovi\ absorption is typically such that $13.2 \la \mlnovi \la 14.4$ and $20 \la \Delta v_{\rm OVI} \la 100$ \kms\ at any $z$ \citep{simcoe02,muzahid12,tripp08}.} This is another indication that the KODIAQ absorbers most likely trace the CGM of actively star-forming galaxies. 

In the previous sections (and see Figs.~\ref{f-comp} and \ref{f-metvsnh1}), we saw that there is a major evolution in the metallicity distribution of the pLLSs and LLSs. There is also a major evolution in the properties of the \ovi\ associated with pLLSs/LLSs. At $z<1$, \ovi\ absorption in the pLLS sample has typically $ 50\la \Delta v_{\rm OVI} \la 150$ \kms\ and  $13.8 \la \mlnovi \la 15$ \citep{fox13}. There is overlap between the low and high $z$ surveys, but broad and strong \ovi\ absorption associated with LLSs and pLLSs at $z<1$ is the exception. Indeed, only two absorbers not captured with these boundary limits shown in Fig.~\ref{f-ovi} are associated each  with a massive, large-scale outflow from a massive star-forming galaxy \citep{tripp11,muzahid15}. These two absorbers  at $z<1$ have $\Delta v \ga 300$ \kms\ and the metallicities of the highly ionized gas and cool ionized gas are similar and super-solar. In the $z<1$ sample, \citet{fox13} also find that strong \ovi\ absorption is not detected when the metallicity of cool gas is low; three of the four broadest and strongest systems at $z<1$ happens to be three of the four most metal-rich pLLSs/LLSs in the sample, with similar properties (although not as extreme) to those of the absorbers associated with  starburst galaxies \citep{tripp11,muzahid15}.  My interpretation of the major difference in the frequency of strong and broad \ovi\ between the low and high $z$ pLLS/LLS surveys is that typically low-$z$ galaxies are much more quiescent than their high redshift counterparts and, indeed, only in the comparatively rare low-$z$ starburst galaxies can we find the extreme \ovi\ absorption seen commonly at high $z$. This also suggests at least for the pLLSs/LLSs with strong \ovi\ absorption that they probe the CGM of some actively star-forming galaxies, not the more diffuse IGM. 

There is therefore little doubt that the strong and broad \ovi\ absorption associated with pLLSs and LLSs typically traces large-scale outflows from very actively star-forming galaxies. The weaker \ovi\ absorbers have likely a wider range of origins, including outflows, inflows, ambient CGM \citep[e.g.,][]{ford14}. It is worth noting that high-$z$ weak \ovi\ absorbers and non-detected \ovi\ absorption are only found in LLSs and SLLSs with $\xh \la -1.5$, i.e., at relatively low metallicity. At $z<1$, \citet{fox13} find that the pLLS bimodal metallicity distribution is reflected to some level in the \ovi\ properties with a lack of low-metallicity pLLSs with strong \ovi\ absorption. The properties of these pLLSs with no detection or of \ovi\ absorption  make them good candidates for cold stream inflows according to recent simulations \citep[e.g.,][]{shen13}. However, a more quantitative and systematic comparison at all $z$ between simulations and observations should be undertaken to gain further insights on these weak \ovi\ absorbers associated with strong \hi\ absorbers. 

\section{Gas accretion via pLLSs and LLSs?}
\label{s-acc}

I have shown that great progress has been made on our knowledge of the class of absorbers known as the pLLSs and LLSs. This success lies in our ability to study sizable and representative samples with good quality (S/N, resolution) spectra at both low and high $z$. With this new knowledge, it is clear that very metal-poor gas is not rare at any $z$ at large overdensities, but it is uniquely probed by the pLLSs and LLSs. This gas is typically not pristine, i.e., it contain some metals, but in some cases down to very low level $\xh \simeq -3.5$. These results provide new stringent empirical results for cosmological hydrodynamical simulations. In particular, there is a strong evolution of the metallicity of the pLLSs/LLSs with $z$, but also a remarkably constant fraction of VMP pLLSs/LLSs over cosmic times. There is also a major shift in the properties of the \ovi\ absorption associated with pLLSs and LLSs, with extremely strong and broad \ovi\ being quite common at high $z$ but rare at low $z$. Since these strong and broad \ovi\ absorbers most likely trace large-scale outflows from actively star-forming galaxies, the dramatic shift of the properties of the \ovi\ absorption associated with pLLSs/LLSs with $z$ certainly captures in part the equivalently dramatic change in the star-formation activity level in galaxies with $z$. 

So are the metal-poor pLLSs/LLSs tracing gas accretion? If we use cosmological simulations to guide us, the answer is very likely yes. Simulations have shown that pLLSs/LLSs may be used to trace cold flows \citep{faucher-giguere11,faucher-giguere15, fumagalli11a, fumagalli14,vandevoort12a, vandevoort12b,hafen16}. Simulated pLLSs and LLSs  at $z\sim 2$--3 or $z<1$ often appear, however, to have too many metals compared to the observations. Only in simulations with very mild stellar feedback \citep{fumagalli11b}, there is some agreement between the observed and simulated metallicity distributions; in this simulation, cold streams are traced mostly by LLSs within 1 or 2 virial radii of galaxies where the gas has only been enriched to $[{\rm X/H}]\simeq -1.8$ with similar scatter to that observed at high or low $z$. However, while mild feedback produces better agreement with the observed metallicity distribution at $z\sim 2$--3, the agreement with the baryon fraction in stars worsens \citep{fumagalli11b}. The FIRE zoom-in simulations at $z<1$ have also recently studied the physical nature of the pLLSs and LLSs \citep{hafen16}. These simulations confirm the general interpretation of the bimodal metallicity distribution at $z<1$: very low metallicity LLSs are predominantly associated with inflows at $z<1$, but higher metallicity LLSs trace gas with roughly equal probability of having recycled outflows (inflows) or outflows. However, the model vs. observation comparison also leads to major differences, including that the simulated metallicity distribution is not bimodal and has a metallicity plateau between about $-1.3$ and $-0.5$ dex at $z<1$. Furthermore, while very low metallicity pLLSs and LLSs are prevalent in the observations, they are not in the FIRE simulations. Possibly recycling and mixing of the gas are too efficient in these simulations to allow for the presence of widespread low-metallicity gas. However, according to all these simulations despite some key disagreements with the observations, there is a consensus that a large fraction of the metal-poor LLSs and pLLSs should probe cold flow accretions onto galaxies, and in fact the covering fraction of the cold flow accretion could peak in the LLS regime. The observations summarized here show that there is a prevalence of metal-poor (and even very metal-poor) pLLSs/LLSs at any $z$, strongly suggesting that they may be tracers of metal-poor gas accretion.

While there are many positive aspects in using the pLLSs and LLSs to study the gas around galaxies, there is a major limitation: it is complicated and often impossible (especially at high redshift) to determine if the gas is actually inflowing into or outflowing from the host galaxy. All the properties so far derived for the pLLSs and LLSs fit with the gas properties found in the simulated cold flow accretion: the \nhi\ range, the low metallicities, the temperatures, densities, length scales, and even some of the galaxy properties \citep[e.g.,][]{ribaudo11,kacprzak12,lehner13}. However, it is unknown from these observations if the gas is actually accreting onto the galaxies. 

The only experiment where we can directly determine the direction of motion along the line-of-sight is the down-the-barrel experiment whereby the galaxy itself is used as the background target: observations of the blueshifted or redshifted absorption relative to the systemic velocity of the galaxy directly inform us if the gas is outflowing or inflowing (e.g., \citealt{heckman00,weiner09,martin12,rubin12}, and see also the chapter by Kate Rubin in this book). In this method, strong metal transitions (like \feii, \mgii) are, however, used, and, hence, it remains ambiguous to know if the gas that is accreting onto a galaxy as recycled gas or from the IGM since the metallicity of the gas is generally unknown. 

\begin{figure}[!t]
\includegraphics[scale=0.5]{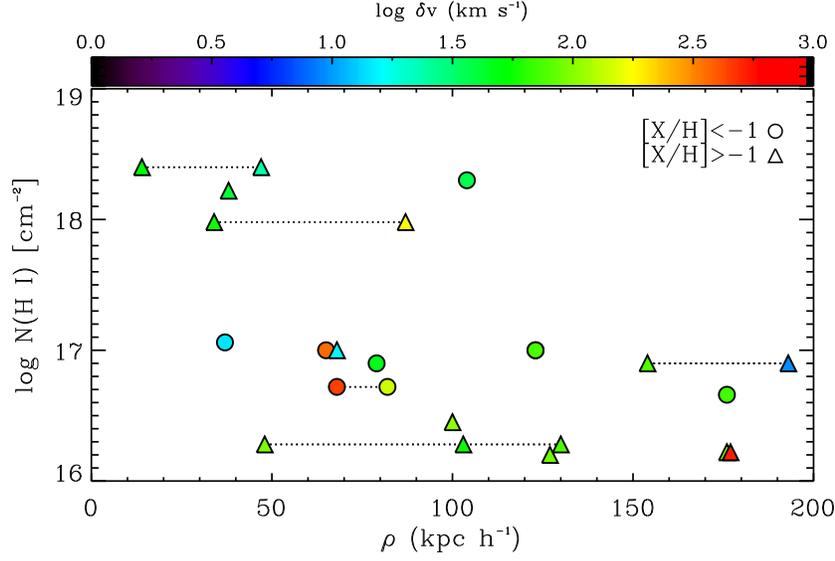}
\caption{The \hi\ column density  as a function of the impact parameter for the pLLSs and LLSs at $z<1$ from the \citet{lehner13} sample. The data are color-coded in velocity separation between the redshift of the absorbers and associated galaxy $\delta v = c |z_{\rm abs} - z_{\rm gal}|/(1+z_{\rm abs})$. Data connected with a dotted line indicate that more than one candidate galaxy was found (note that the triangles nearly overlapping at $\mlnhi \simeq 16.2$ and $\rho \simeq 176$ kpc are also a pair of candidate galaxies for the same absorber).  }
\label{f-gal}       
\end{figure}

Another method for indirectly determining the origin of the gas has been to characterize geometry of the galaxy relative to the QSO line of sight (see for a full description of this method the chapter in this book by Glenn Kacprzak). Several \mgii\ surveys have demonstrated that a class of strong \mgii\ absorbers at small impact parameters ($\rho\lesssim40$ kpc) are primarily found along the galaxies' minor axes, consistent with an origin from large-scale outflows \citep[e.g.,][]{bordoloi11, kacprzak12b,bouche12}. However, \citet{kacprzak12} has also shown one example of a very metal-poor LLS observed projected along the minor axis of the associated galaxy albeit at a projected distance of 104 kpc. Evidently all absorption detected in the spectra of QSOs probing gas along the projected minor axis of galaxies may not be produced by winds, especially at these large impact parameters. Metal-poor gas may also not necessarily align with the projected major axis of the galaxy as expected from some simulations \citep[e.g.,][]{stewart11}, i.e., other mechanisms than cold flow accretion might be at play, as, e.g., the precipitation model proposed by \citet{voit15}. 

One can also undertake a combination of both methods just described above whereby absorption is observed toward a galaxy (down-the-barrel) and toward a QSO transversely projected at some distance \citep[e.g.,][]{kacprzak14,bouche16}. I suspect the number of galaxy-QSO pairs will increase dramatically in the future with IFU observations and larger aperture telescope.

Cosmological simulations  show a preference for inflowing streams to align with the major axis of galaxies, metal-enriched outflows along the minor axis, and tidal debris more uniformly distributed (e.g., \citealt{stewart11,vandevoort12a,shen13}). This combined with the empirical results from  \mgii\ absorber-galaxy survey strongly motivated the Bimodal Absorption System Imaging Campaign (BASIC; PI: Lehner), which aimed to jointly determined the correlation (if any) between the geometry of the galaxies with the metallicities of the pLLSs/LLSs. This program combines newly acquired and archival ACS images as well as ground-based galaxy imaging and spectroscopy (LBT/MODS-LBC; MUSE; Keck LRIS) of all the fields from the \citet{lehner13} sample. With the ongoing BASIC survey, we will measure the geometry,  morphology, environment of galaxies associated with $\sim$30 pLLSs/LLSs and in particular test if the bimodal metallicity distribution is connected to the geometric orientation of galaxies as predicted by simulations. 

While it is too early to report on any specific results from BASIC, in Fig.~\ref{f-gal}, I present an updated figure from \citet{lehner13} showing \nhi\  as a function of the impact parameter $\rho$ of the possible associated galaxies at $z<1$. This figure differs from Fig.~9 in \citet{lehner13} in several important aspects. First, in \citet{lehner13} only the ``closest'' galaxy in redshift and impact parameter to a given absorber was considered, while here I show all the potential candidate galaxies for a given absorber (not just one). Second, I differentiate the low and high metallicity absorbers. Third, the velocity separation $\delta v = c |z_{\rm abs} - z_{\rm gal}|/(1+z_{\rm abs})$ between the redshift of the absorbers and associated galaxy is indicated. I stress that this is still very preliminary since not the entire sample has been obtained and  the galaxy completeness has not been systematically assessed yet. However, as \citet{lehner13} indicated at $z<1$ for 11 LLSs, now for 16 pLLSs/LLSs (7 metal-poor absorbers, 9 metal-rich absorbers) where their fields have been observed sufficiently deeply ($\ga 0.1 L*$), at least one galaxy is found within $\rho \la 200$ kpc and typically $\delta v \la 100$ \kms\  (except for a few cases, see Fig.~\ref{f-gal}). From Fig.~\ref{f-gal}, there is no hint of a  difference between low and high metallicity absorbers with impact parameters. With the current sample, there is, however, a strong hint that $\xh \ge -1$ pLLSs/LLSs are found associated with more than one galaxy, while $\xh < -1$ pLLSs/LLSs are all but one associated with a single galaxy.  Further studies will be needed to interpret the galaxy/LLS connection, but it is evident that we will learn soon more about the environments and hence the nature of these systems. With the current data, it is reasonable to conclude that the VMP pLLSs/LLSs at $z<1$ are indeed probed of the extended of the CGM of galaxies (not necessarily of a single galaxy at least for metal-rich absorbers). 

At $z>2$, the geometry of the galaxies will be difficult to decipher, but observations with the VLT/MUSE or Keck/KCWI of the fields of view of the pLLSs and LLSs  will shed light on the galaxy environments of the pLLSs and LLSs. These observational efforts are in their infancy, but recent MUSE observations show that the connection between very metal-poor LLSs and star-forming galaxies fed by cold streams is quite plausible at high $z$ \citep{fumagalli16a} (at least for these VMP LLSs). 

\section{Conclusions and future directions}
\label{s-summ}
As a massive reservoir of galactic baryons and metals, the mediator of galactic feedback, and the fuel for star formation, the CGM plays a major role in galaxy evolution, yet the full nature of that role and how it evolves with cosmic time remains to be determined. With the pLLS and LLS surveys at $z<1$ and $z>2$, we have made critical advances on our understanding of the CGM of galaxies over the last few years. It is obvious from these studies that pLLSs and LLSs trace different processes and have several origins. It is also quite obvious now that, at the high overdensities probed by pLLSs and higher column density systems, the gas is very rarely pristine. There is a large fraction of pLLSs and LLSs that have metallicities similar to those observed in SLLSs and DLAs. However, there is also a substantial fraction (20\%--45\%)  of very metal-poor gas   at all studied $z$ intervals. At $z<1$, there is also little doubt that the pLLSs and LLSs are tracing the dense regions of the CGM of galaxies, while at higher $z$ the question is still open (although pLLSs or LLSs associated with strong \ovi\ absorption are quite likely to probe the CGM of some very actively star-forming galaxies). At $z<1$, these findings indicate that galaxies have large reservoir of metal-poor gas at projected physical distances $\la 100$--200 kpc, which may eventually feed the galaxies, but the exact mechanisms on how this occurs (and if this really occurs) are still an open question.

Hence as other techniques to study the gas around galaxies, the evidence for large-scale outflows is quite clear and unambiguous (pLLSs and LLSs with strong and broad \ovi\ absorption trace large-scale outflows from starburst galaxies), but the evidence for accretion is still not as equivocal. A good way to conclude this chapter on LLSs and accretion is by quoting Michele Fumagalli  from his talk at the IGM@50 {\it Is the Intergalactic Medium Driving Star Formation}\ Spineto conference last year: ``So far, there is no [major] empirical evidence against the association between [pLLS]/LLSs and cold accretion as put forward by theory". I would add, however, that we should not dismiss yet other accreting gas models that might be at play \citep[e.g.,][]{marasco13,voit15}. \\

I predict the next five years are going to be equally as rich in new discoveries as the previous five years if not more:
\begin{itemize}
\item At all $z$, samples where we will be able to robustly study the metallicity distribution will increase by a factor 2--4 at $z<1$ ($\ga 100$ absorbers) and a factor $>5$--10 at $z>1.5$ ($>150$--$300$ absorbers) thanks to the growing archives at MAST, KOA, and ESO. Not only the samples will increase, but there will also be better sampling in redshift space (e.g., finer redshift sampling will be possible at $2<z<4$ as well as in the redshift interval $1<z<2$ where currently there is no information) and in \nhi\ (e.g, the LLS sample will increase at $z<1$ as well as the weak pLLS sample at the transition of the \lya\ forest at both $z<1$ and $z>2$). 
\item HST images and integral field spectroscopy with moderate to high spectral resolution will transform our knowledge of the LLS and pLLS galaxy environments. We are now in position to study the galaxies associated with absorbers where some of their key properties are determined (the metallicity of course, but also their various gas phases -- including the highly ionized probed by \ovi, kinematics, etc.). The recent and ongoing advent of IFU spectroscopy (and future AO) systems on ground-based 8--10m class telescopes (MUSE, KCWI) will speed up the acquisition of the galaxies, but more critically galaxies at $z\la 1$ can be spatially and kinematically resolved providing a new wealth of information extremely difficult to acquire in the past \citep[see, e.g.,][]{fumagalli14a,fumagalli16a,bouche16,peroux16}. As these efforts will be undertaken at both low and high $z$, these galaxy surveys will enable the study of their evolution over much of the age of the universe.  On a longer time frame, extremely large telescopes, with their huge light gathering power coupled with diffraction-limited angular resolution, will even further revolutionize our understanding of the CGM of galaxies at both high and low $z$.
\item Finally, observers and simulators are working more and more together and I expect from these collaborations more insights will be gained on the nature of the gas probed by the pLLSs and LLSs. In particular, new tools are being developed so that results from the observations and simulations can be more directly comparable and more easily interpretable (e.g., by treating metallicities of individual absorbers as posterior probability distribution functions, see \citealt{fumagalli16}). 
\end{itemize}

\begin{acknowledgement}
It is a pleasure to thank all my collaborators, current and past graduate students who all have been key on shedding light on these (now much less mysterious) absorbers over the last few years, and, in particular, Chris Howk, John O'Meara, and Xavier Prochaska who have been critical players on many aspects to push forward these projects at both low and high redshifts. I also thank Chris Howk for reading and providing useful comments on this manuscript and Lorrie Straka for reducing the MUSE data and providing the galaxy redshifts that help to make Figure~\ref{f-gal}.  The writing and some of the research presented in this work has been supported by NASA through the Astrophysics Data Analysis Program (ADAP) grant NNX16AF52G, HST-AR-12854 and HST-GO-14269 from the Space Telescope Science Institute, which is operated by the Association of Universities for Research in Astronomy, Incorporated, under NASA contract NAS5-26555.
\end{acknowledgement}

\end{document}